\documentclass[3p,times]{elsarticle}

\usepackage{ecrc}
\usepackage{bm}

\volume{00}

\firstpage{1}

\journalname{Nuclear Physics A}

\runauth{U. Heinz {\it et al.}}

\jid{npa}

\jnltitlelogo{Nuclear Physics A}

\usepackage{amssymb}
\usepackage{mathrsfs}
\usepackage{amsfonts}
\usepackage{amsthm}
\usepackage{latexsym}
\usepackage{slashed}

\biboptions{square,comma,numbers,sort&compress}

\usepackage[figuresright]{rotating}

\usepackage{color}
\usepackage{hyperref}

\newcommand{\Blue}[1]{\textcolor{blue}{\bf{#1}}}

\begin{document}

\begin{frontmatter}



\dochead{}

\title{Electromagnetic fingerprints of the Little Bang}

\author{Ulrich Heinz\footnote{Invited speaker. Email address: heinz@mps.ohio-state.edu}}
\author{Jia Liu}
\author{Chun Shen}

\address{Department of Physics, The Ohio State University, Columbus, Ohio 43210-1117, USA}

\begin{abstract}
Measurements of thermal photons emitted from the rapidly expanding hot and dense medium ("Little Bang") formed in ultra relativistic heavy-ion collisions, and their current theoretical interpretation, are reviewed.
\end{abstract}


\end{frontmatter}


\section{Introduction}
\label{intro}

Direct\footnote{``Direct'' means that they do not arise from the electromagnetic decay 
     of long-lived but unstable hadrons after their final decoupling, 
     but are emitted directly from the medium.}
photons\footnote{We will here focus on real photons; for a discussion of virtual photons decaying into lepton-antilepton pairs (``dileptons''), see contributions by J. Butterworth, M. Laine, A. Uras, and G. Vujanovic elsewhere in this volume.}
have long been recognized as clean penetrating probes of the dense, strongly interacting medium created in relativistic heavy-ion collisions. {\em Thermal} photons are a subclass of direct photons that are radiated by a medium in (approximate) local thermal (but not necessarily chemical) equilibrium. They are valuable messengers from the evolving fireball (``Little Bang'') formed in the collision, and they diagnose the types of matter (glasma, quark-gluon plasma, hadron resonance gas) which make up this fireball during different stages of its evolution. Superimposed on the thermal signal are ``prompt'' contributions from hard scatterings between the incoming partonic constituents of the incoming nuclei and from jet fragmentation \cite{Owens:1986mp}, from jets interacting with the thermal medium \cite{Fries:2002kt,Zakharov:2004bi}, as well as pre-equilibrium photons from the glasma \cite{McLerran:2014hza} and from secondary scatterings between hard partons before the medium thermalizes \cite{Bass:2002pm}. Thermal photons tend to dominate the direct photon energy spectrum below about $2.5$\,GeV at RHIC and below about $3$\,GeV at the LHC \cite{Turbide:2007mi,Chatterjee:2013naa,Shen:2013vja}. Prompt and pre-equilibrium photons, except those from the glasma \cite{McLerran:2014hza}, contribute at higher photon energies where they eventually overwhelm the thermal radiation.

First predictions for thermal photons radiated from the Little Bang date back more than two decades (see e.g. \cite{Ruuskanen:1992hh} and references therein), but it took until 2008 for the first unambiguous measurement of an excess over pQCD expectations for prompt photon emission, by PHENIX in $200\,A$\,GeV Au+Au collisions \cite{Adare:2008ab}, which was later confirmed by ALICE in 2012 in 2.76\,$A$\,TeV Pb+Pb collisions at the LHC \cite{Wilde:2012wc}. The shape of the excess can be fit with a simple exponential in the photon transverse momentum $p_T$, with an inverse slope $T_\mathrm{eff}{\,=\,}221{\,\pm\,}19{\,\pm\,}19$\,MeV for central (0-20\% centrality) Au+Au collisions at RHIC \cite{Adare:2008ab} and $T_\mathrm{eff}{\,=\,}304{\,\pm\,}51$\,MeV for semi-central (0-40\% centrality) Pb+Pb collisions at the LHC \cite{Wilde:2012wc}. While both values lie above the critical temperature of about 155-165 MeV for quark-gluon plasma formation, we show next that the correct interpretation of these slope parameters is much more involved \cite{Shen:2013vja}.

\section{Photon spectra and their inverse slopes}
\label{spectra}

Since photons are emitted without re-interaction from all stages of the Little Bang, the measured photon signal represents a space-time integral of the local thermal emission rate $R{\,=\,}dN^\gamma/d^4x$, boosted to the lab frame by the collective expansion velocity, over the four-volumes corresponding to each value of the fireball temperature: $E (dN^\gamma/d^3p) = \int d^4x\, \left[ E \bigl(dR^\gamma/d^3p\bigr)(p{\cdot}{u}/T)\right]_{T(x),p{\cdot}{u}(x)}$ (where $u^\mu(x)$ is the flow four-velocity in the lab frame of the fluid cell at point $x$ with local temperature $T(x)$, and $\bar{E}{\,=\,}p{\cdot}u(x)$ is the local rest frame energy of a photon carrying momentum $p^\mu{\,\equiv\,}(E,\bm{p}_\perp,p_L)$ in the lab frame.).      

%
\begin{figure*}[htb]
\begin{center}
\begin{tabular}{cc}
    \includegraphics[width=0.49\linewidth,clip=]{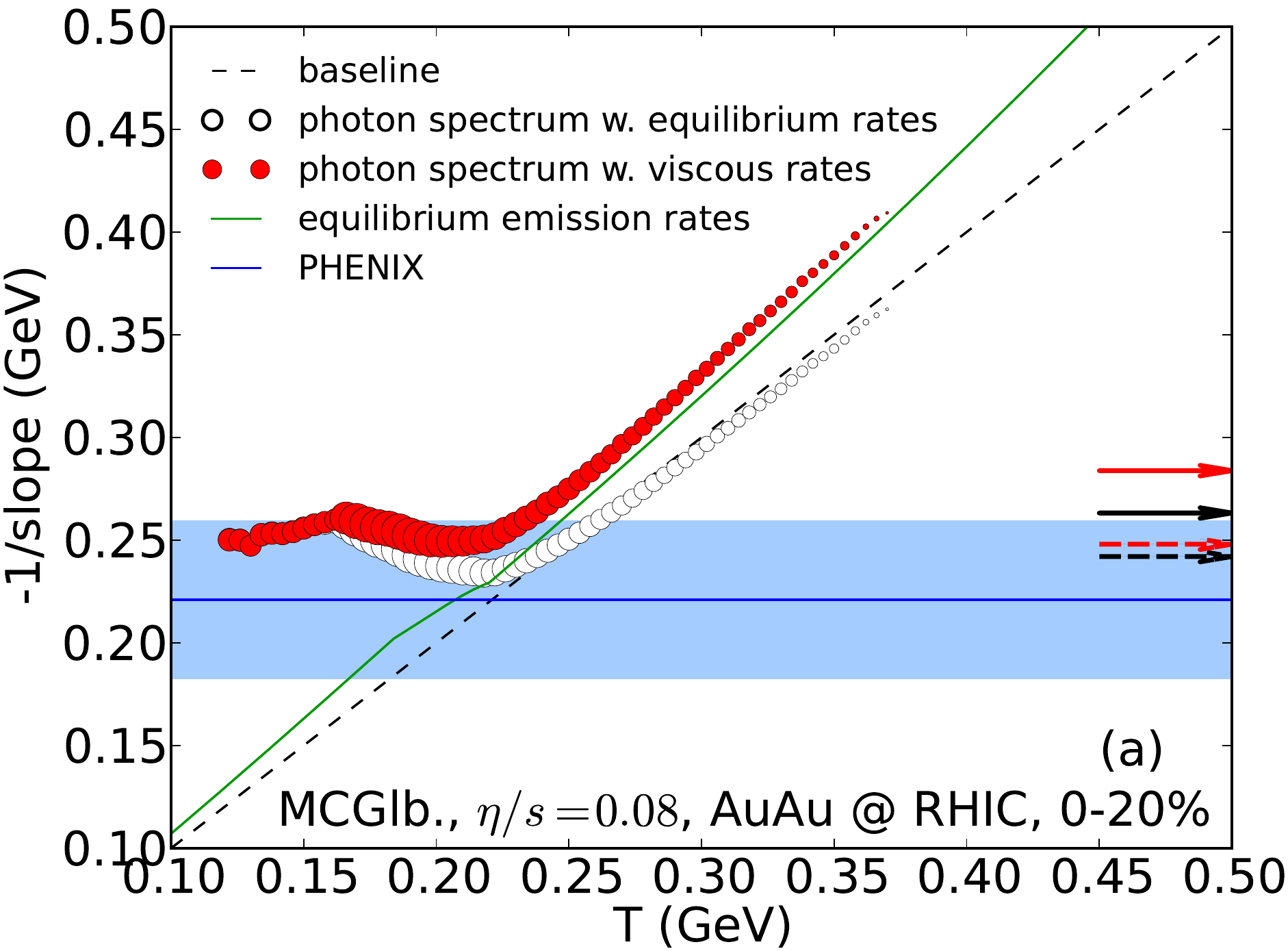}
    \includegraphics[width=0.49\linewidth,clip=]{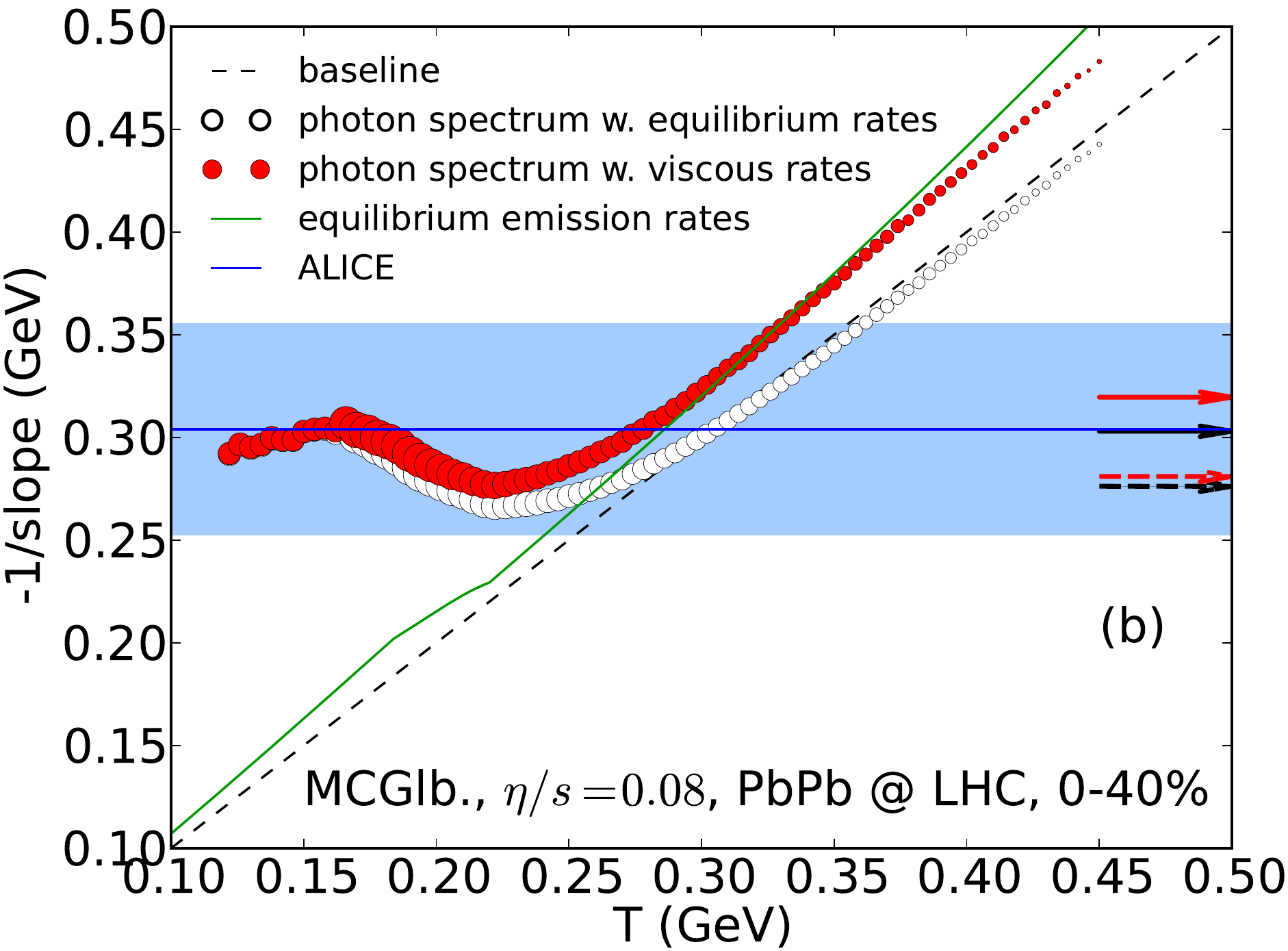}
\end{tabular}
\end{center}
    \vspace*{-3mm}
\caption{(Color online)
    The inverse photon slope parameter $T_\mathrm{eff}{\,=\,}{-}1/\mathrm{slope}$ as a function 
    of the local fluid cell temperature, from the equilibrium thermal emission rates (solid green
    lines) and from hydrodynamic simulations (open and filled circles), compared with the 
    experimental values (horizontal lines and error bands), for (a) Au+Au collisions at RHIC 
    and (b) Pb+Pb collisions at the LHC. The experimental values and error bands are from 
    the PHENIX Collaboration \cite{Adare:2008ab} in (a) and from the ALICE Collaboration 
    \cite{Wilde:2012wc} in (b). We note that the corresponding plot for Au+Au collisions at $20{-}40\%$ 
    centrality looks very similar to the case of $0{-}20\%$ centrality shown in panel (a), in agreement
    with what the PHENIX Collaboration \cite{Adare:2008ab} found. The horizontal arrows indicate 
    the inverse slope of the space-time integrated thermal photon spectrum: solid black and red 
    arrows correspond to the full calculation using thermal equilibrium and viscous-corrected rates,
    respectively; dashed black and red arrows correspond to the analogous spectra obtained by 
    shutting off (by hand) all radiation from $T{\,>\,}250$\,MeV, to estimate potential effects from 
    delayed quark production in an initially gluon-dominated glasma \cite{Shen:2013vja}. The areas 
    of the bubbles are proportional to the total photon yield they represent.   
    \label{F1}
    }
\end{figure*}
%

Figure~\ref{F1} \cite{Shen:2013vja} illustrates the connection between the local rest frame temperature of the emitting volume elements and the inverse slope of the photon $p_T$-spectrum emitted from those elements as measured in the lab frame, for (2+1)-dimensional (longitudinally boost-invariant) viscous hydrodynamic simulations at RHIC (a) and LHC energies (b) \cite{Shen:2013vja}. Even in the local rest frame, the inverse slope of the photon spectrum (green solid line) is larger than the temperature, due to a logarithmic factor in the matrix element that favors radiation at higher energies: $E dR^\gamma/d^3p \sim \exp(-E/T)\ln(E/T)$ \cite{Kapusta:1991qp,Baier:1991em}. An isotropic fireball of temperature $T$ that is at rest in the lab frame emits photons at rapidity $y$ with an intensity and inverse slope controlled by $T_\mathrm{eff}{\,=\,}T/\cosh{y}{\,<\,}T$ ; a boost-invariant superposition of fireballs with temperature $T$ thus emits photons at mid-rapidity in the lab with an inverse slope parameter that is smaller than in the local rest frame. This explains why in Fig.~\ref{F1} for high temperatures (i.e. for early times when longitudinal expansion dominates) the white bubbles (corresponding to emission from a viscously expanding fireball with {\em thermal equilibrium emission rates}) lie significantly below the green line, and even slightly below the dashed black line reflecting the true emission temperature $T$. However, in a viscously expanding system photons are not emitted with {\em thermal equilibrium emission}: for anisotropic expansion, shear viscosity causes the local rest frame momentum distribution to become anisotropic, too. For heavy-ion collisions which, at least initially, expand much more rapidly in longitudinal than transverse direction, this leads to a local momentum distribution that is broader in transverse than longitudinal direction. This viscous correction increases the inverse slope of the emitted photons, as reflected in the red bubbles in Fig.~\ref{F1} compared to the white ones. These viscous corrections to the photon emission rate are quite large during the earliest stage of the Little Bang but become negligible in its late hadronic stage.

Most important for the interpretation of the experimentally observed inverse slope is, however, the effect of radial flow, seen to kick in when the fireball temperature drops below ${\sim\,}250$\,MeV at RHIC and below ${\sim\,}280$\,MeV at the LHC. The blueshift of the photon spectrum caused by radial flow boosts the inverse slope to values significantly above the true emission temperature, to the extent that below $T{\,\sim\,}220$\,MeV the inverse photon slope at both RHIC and LHC {\em increases} again while $T$ continues to drop, until the system hits the chemical freeze-out temperature at 165\,MeV below which the equation of state changes, leading to faster cooling and causing the inverse slope to decrease again. From the size of the bubbles one infers that more than 2/3 of the finally observed thermal photon yield is emitted with inverse slopes around 240\,MeV at RHIC and around 280\,MeV at the LHC, even though the true temperatures at which those photons were emitted were significantly smaller.       

The message from Fig.~\ref{F1} can be summarized as follows: The measured yield of thermal photons, and the slope of their $p_T$-spectrum, is dominated by fireball temperatures below 250 MeV that are strongly blue-shifted by radial flow. A large fraction of those photons comes from the transition region between QGP and hadron gas, with smaller fractions contributed by the very hot early QGP and the cooler late hadronic stages. Evidence for a very dense initial state exists, but it comes only indirectly from the photon spectra but more directly from the observed strong collective flow which requires large initial pressure gradients to build up and which is also seen in all the hadron spectra.
 
\section{Thermal photon anisotropic flow}
\label{flow}
 
The first prediction of thermal photon elliptic flow in 2005 \cite{Chatterjee:2005de} was based on ideal fluid %
\begin{figure}
\begin{center}
\begin{tabular}{cc}
    \includegraphics[width=0.85\linewidth]{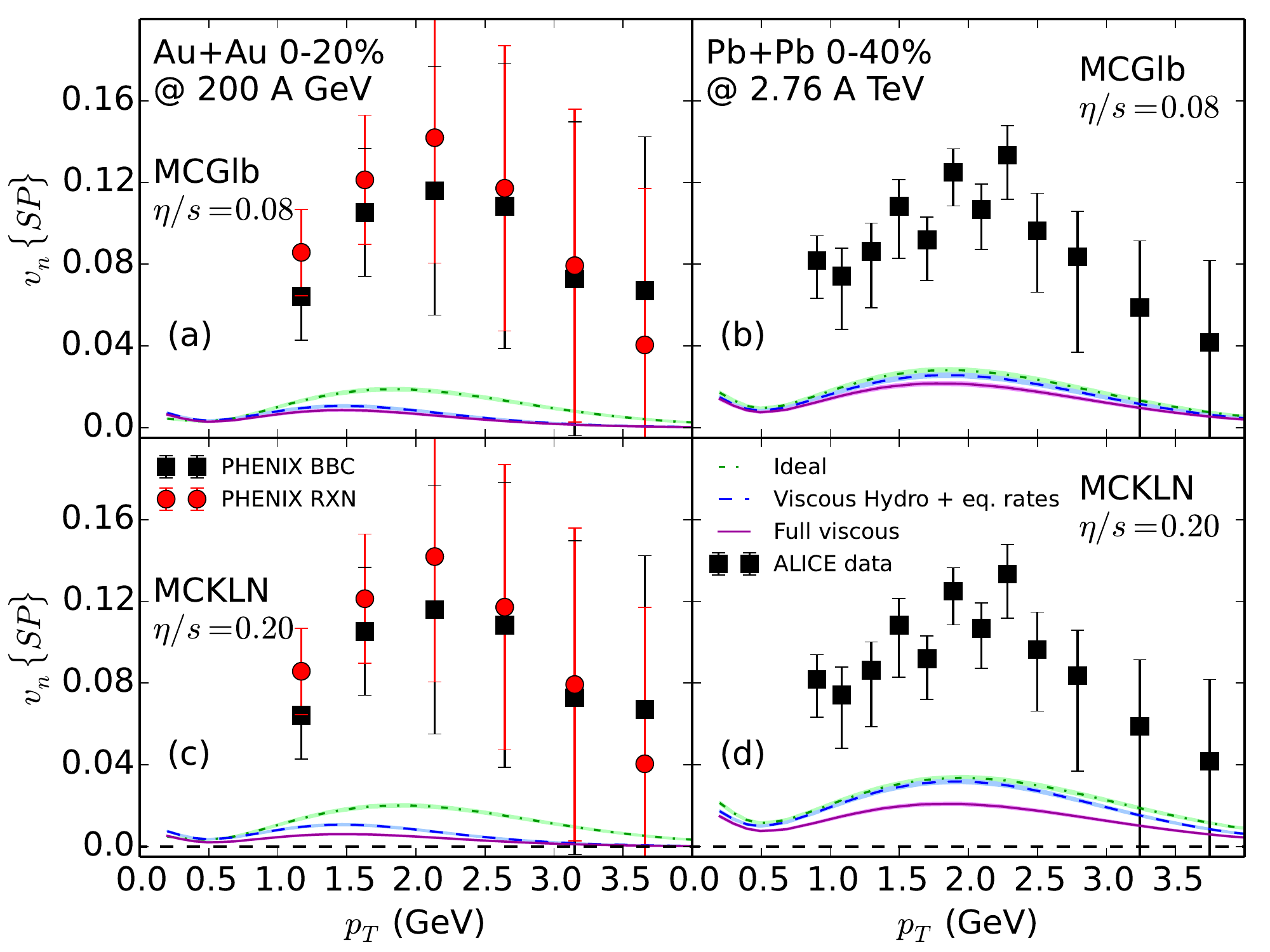}
\end{tabular}
\end{center}
    \vspace*{-5mm}
\caption{(Color online)
    Direct (i.e. thermal + prompt) photon elliptic flow as a function of $p_T$, for 200\,$A$\,GeV (0-20\%)
    Au+Au \cite{Adare:2011zr} (a,c) and 2.76\,$A$\,TeV (0-40\%) Pb+Pb collisions (b,d) 
    \cite{Lohner:2012ct}. Lines show 
    predictions from event-by-event (2+1)-d viscous fluid dynamics with MC-Glauber (a,b) and 
    MC-KLN (c,d) initial conditions, propagated with shear viscosities $\eta/s{\,=\,}0.08$ and 0.2, 
    respectively, as required for reproduction of the charged hadron elliptic flow in these collisions. 
    Green dotted lines: ideal hydro with thermal equilibrium emission rates; blue dashed lines: 
    viscous hydro with thermal equilibrium emission rates; solid magenta lines: viscous hydro with 
    viscous-corrected emission rates. 
    \label{F2}
    }
\end{figure}
%
dynamics with smooth, event-averaged initial conditions and a chemical equilibrium equation of state with a first-order phase transition. Modern state-of-the-art hydrodynamic calculations of direct photons from the Little Bang use event-by-event simulations of viscous fluid dynamics with Monte Carlo sampled fluctuating initial density profiles that evolve with a lattice-QCD-based equation of state with chemical decoupling around $T_\mathrm{chem}{\,\sim\,}165$\,MeV \cite{Qiu:2011hf,Dion:2011pp}.\footnote{An automated code package implementing this workflow is available at the URL \Blue{\href{https://github.com/chunshen1987/iEBE.git}{https://github.com/chunshen1987/iEBE.git}}.}
Other approaches are based on a solution of kinetic equations \cite{Linnyk:2013hta}, or couple the hydrodynamic simulation of the QGP with a microscopic simulation of the late hadron resonance gas stage (this ``hybrid'' approach has so far only been applied to hadron spectra, but not yet to photon emission). Work on including photons from a pre-equilibrium stage \cite{Gale:2012rq} before the onset of hydrodynamics is ongoing.

First experimental data on direct photon elliptic flow came in 2011 from a study of 200\,$A$\,GeV Au+Au collisions at RHIC by the PHENIX collaboration \cite{Adare:2011zr}, closely followed by ALICE which presented photon elliptic flow measurements in 2.76\,$A$\,TeV Pb+Pb collisions at the LHC at {\sl Quark Matter 2012} \cite{Lohner:2012ct}. The surprise from both measurements was that the direct photon elliptic flow was large, almost as large as that of charged hadrons, peaking at values around 10-12\% near $p_T{\,=\,}2$\,GeV/$c$. This contradicted ideal fluid dynamic predictions \cite{Chatterjee:2005de,Holopainen:2011pd,Chatterjee:2013naa} which for thermal photons gave $v_2(p_T)$ signals that peaked around 3-4\%, dropping below 2\% once prompt photons (which carry zero elliptic flow) were included in the direct photon signal. This embarrassing disagreement became since known as the ``direct photon flow puzzle''. 

Viscous effects affect the hydrodynamic predictions for photon flow anisotropies in three places: (1) To compensate for viscous heating and maintain a constant final charged hadron multiplicity density, the Little Bang must start with a smaller initial temperature in the viscous case than in an ideal fluid simulation. This reduces the rate of photon emission at early times where radial, elliptic and higher-order anisotropic flows are weak, and thus {\em increases} the overall direct photon elliptic flow, by increasing the relative weight from the late collision stages where radial and anisotropic flows are strong. (2) Shear viscosity inhibits the conversion of anisotropies in the initial pressure gradients into hydrodynamic flow anisotropies. This {\em reduces} elliptic and anisotropic flows of all particle species. (3)~In an anisotropically expanding system, such as the Little Bang which expands much faster along the beam direction than transverse to it, shear flow, through shear viscous effects, distorts the local rest frame momentum distribution, making it anisotropic. This affects the photon emission rate, leading to a strong {\em reduction} of photon transverse flow anisotropies, especially at early times where this longitudinal-transverse flow shear is strongest \cite{Song:2007ux,Dion:2011pp,Shen:2013cca}. Of these three mechanisms, the third has the strongest influence on the anisotropic flow coefficients. Their combined effect is a significant reduction of the photon elliptic and higher-order anisotropic flows compared to the earlier ideal-fluid simulations \cite{Chatterjee:2005de,Holopainen:2011pd,Chatterjee:2013naa}. Obviously, this exacerbates the disagreement between theoretical predictions and experimental data, as shown in Fig.~\ref{F2}.

%
\begin{figure}
\begin{center}
\begin{minipage}{0.4\linewidth}
    \includegraphics[width=\linewidth,clip=]{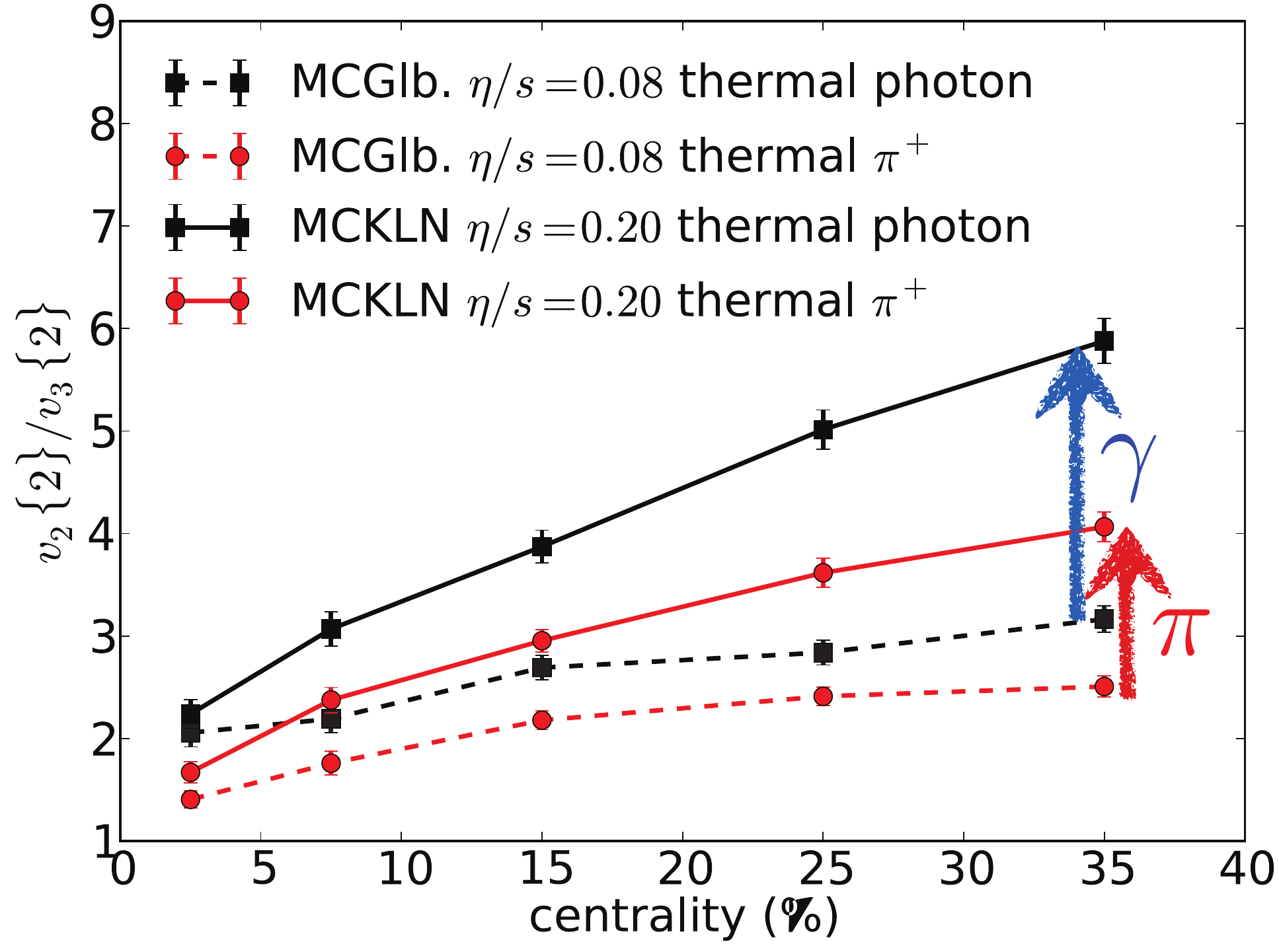}
\end{minipage}
\hspace*{7mm}
\begin{minipage}{0.3\linewidth}  
\vspace*{-8mm}  
\caption{(Color online)
    The elliptic-to-triangular flow ratio for thermal photons and pions as a function of collision centrality,
    for 0-40\% 2.76\,$A$\,TeV Pb+Pb collisions at the LHC, for two different shear viscosities and initial
    conditions that give the same charged hadron $v_2$. Here the ratio is plotted using the rms flows 
    $v_n\{2\}{\,\equiv\,}\langle v_n^2\rangle^{1/2}$ -- the analogous ratio for the more easily measured
    (but slightly more difficult to interpret) scalar product $v_n$ (see C. Shen's contribution to this 
    volume) is found in \cite{Shen:2013cca}.
    \label{F3}
    }
\end{minipage}
\end{center}
\vspace*{-5mm}
\end{figure}
%

Event-by-event hydrodynamics of fluctuating initial conditions allows to compute also the higher-order flow harmonics for direct photons \cite{Shen:2013cca}. In this context the triangular flow of photons plays a special role \cite{Shen:2013cca}: While hydrodynamics predicts in semi-central Au+Au or Pb+Pb collisions elliptic and triangular flows of similar magnitudes ($v_2{\,\sim\,}(2{-}3)v_3$, see Fig.~\ref{F3}), other more exotic mechanisms \cite{Basar:2012bp}, invoking an anomalous coupling in non-central collisions to the magnetic field perpendicular to the reaction plane to explain the large measured photon elliptic flow, are not expected to yield an appreciable triangular flow. The ratio $v_2/v_3$ can thus distinguish between the hydrodynamic and anomalous scenarios. Note that this ratio is insensitive to contributions to the photon yield that carry no or only very small anisotropic flows, such as prompt and pre-equilibrium photons \cite{Shen:2013cca}. In Fig.~\ref{F3} we compare this ratio for thermal photons and pions. One sees that both its centrality dependence and sensitivity to shear viscosity are stronger for photons than for pions, reflecting the on average earlier emission of the photons. In combination with hadrons, thermal photons are therefore a useful viscometer for the quark-gluon plasma stage in the Little Bang \cite{Dusling:2009bc,Shen:2013cca}.

\section{Paths towards resolving the direct-photon flow puzzle}
\label{paths}

\begin{figure}[htb]
\begin{center}
    \includegraphics[width=0.95\linewidth]{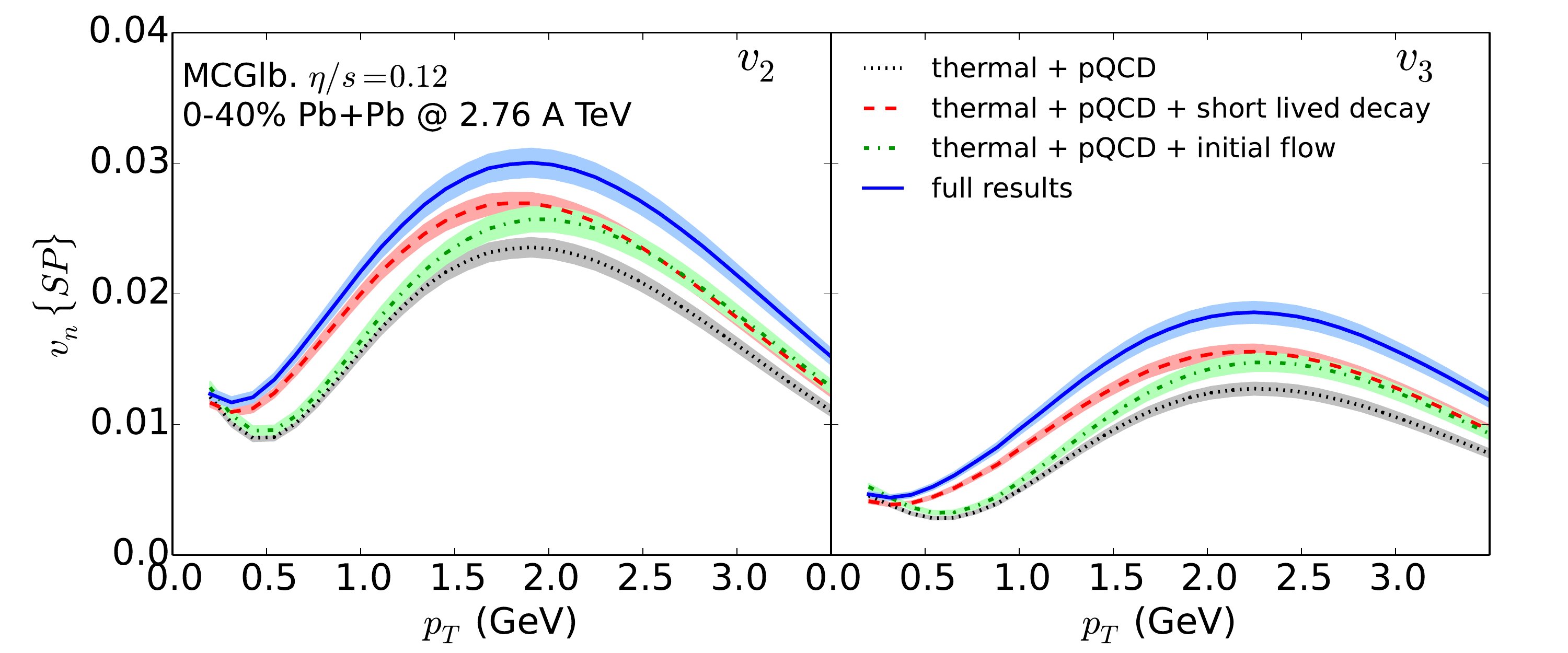}
\end{center}
    \vspace*{-5mm}
\caption{(Color online)
    Effect of pre-equilibrium flow and photons from the decay of short-lived hadronic resonances after
    hadronic decoupling on the direct (thermal + prompt) photon elliptic and triangular flow. Shaded 
    regions indicate statistical uncertainties related to event-sample size.
    \label{F4}
    }
\end{figure}

We saw that hydrodynamic simulations seriously underpredict the measured photon elliptic flow. They also significantly underpredict the photon yield at low $p_T{\,<\,}2$\,GeV \cite{Turbide:2007mi,Chatterjee:2013naa,Shen:2013vja}. Inclusion of pre-equilibrium photons may help to boost the photon yield, but at the expense of further diluting the photon elliptic flow which is small in the pre-equilibrium stage (unless the pre-equilibrium dynamics somehow delays the entire QGP formation process and its hydrodynamic cooling until significant elliptic flow has been generated \cite{McLerran:2014hza}). Pre-equilibrium dynamics has, however, the potential to increase the photon elliptic flow after thermalization. In contrast to hadrons, which are only sensitive to the final elliptic flow at hadronic freeze-out, even a small increase of the flow anisotropy will affect the elliptic flow of photons {\em emitted during all stages} of the collision; given the small flow predicted so far, this could turn out to be a significant improvement. Comparing the green and black curves in Fig.~\ref{F4} illustrates the possible effect of pre-equilibrium flow: while the black curves assume zero transverse flow at the beginning of the hydrodynamic stage at $\tau_0{\,=\,}0.6$\,fm/$c$, the green dot-dashed curves allow the massless partons created at $\tau{\,=\,}0$ with a bumpy initial density profile to free-stream until $\tau_0{\,=\,}0.6$\,fm/$c$, where the resulting energy momentum tensor is Landau-matched to viscous hydrodynamic form, yielding an non-zero anisotropic initial flow profile (together with a non-zero initial viscous pressure tensor). Since massless partons move with light speed, free-streaming likely overestimates the initial flow at the matching time somewhat. Still, its effect on the photon anisotropic flow is not negligible, and any effect that helps to increase the elliptic flow signal is welcome.       

%
%

Mostly we seem, however, to be missing strength in the photon emission at late times. The large elliptic flow that must be carried by these missing photons, the centrality dependence of their yield (see B.\ Bannier's contribution to this volume and hydrodynamic calculations presented in \cite{Shen:2013vja}), and the fact that we lack thermal photons in the spectrum mostly at low $p_T$ \cite{Shen:2013vja,Chatterjee:2013naa}, all indicate that most of the missing photons must come from either the phase transition region or the hadronic rescattering phase. This is supported by findings that models that are able to reproduce the measured direct photon spectra and elliptic flow better than hydrodynamics either have a larger space-time volume contributing to low-temperature emission, such as the parametrized expanding fireball model of Rapp and van Hees \cite{vanHees:2007th}, or continue to radiate even at energy densities well below the typically implemented hadronic freeze-out densities in hydrodynamic codes that were tuned to describe the hadron spectra, due to ongoing hadronic collisions that, although rare, can last for a long time in a hadronic cascade \cite{Linnyk:2013hta}. Also, some of the many hadronic resonances emitted from the decoupling hadron resonance gas have photon-producing decay channels that were not included in hydrodynamic simulations before (we thank R. Rapp for pointing this out). These resonances decouple with large anisotropic flows and pass them on to their decay-photon daughters. Even though the corresponding branching ratio for each resonance is small, after adding up photons from all such short-lived resonance decays, the direct photon elliptic and triangular flows are found to increase by up to 25\%. This is shown by the difference between the red dashed and black dotted curves in Fig.~\ref{F4}. Adding both the pre-equilibrium flow and post-freeze-out resonance decay contributions thus enhances the photon elliptic and triangular flows by 40-50\% (blue solid lines in Fig.~\ref{F4}).

Finally, the present hydrodynamic models do not include all the same hadronic photon-production channels as the parametrized fireball \cite{vanHees:2007th} and microscopic scattering \cite{Linnyk:2013hta} models. In particular, the hadronic rates implemented in viscous hydrodynamics \cite{Dion:2011pp,Shen:2013vja,Shen:2013cca} include only mesonic 2-to-2 scattering processes, leaving out baryon-induced medium modifications of the vector and axial vector meson spectral functions \cite{Turbide:2003si} as well as bremsstrahlung emitted in collisions with mesons and baryons \cite{Linnyk:2013hta} because we do not know yet how to correct these rates for viscous effects. Such contributions can increase the hadronic emissivity by up to a factor 4 at low $p_T$ \cite{Linnyk:2013hta}. The switching between QGP and hadronic emission rates, as well as a possible boost of both types of emission rates near the phase transition due to critical scattering (Ralf Rapp, private communication), remain issues that require additional investigation. 

At this point, no single large source of missing photons has been identified that would promise an instantaneous resolution of the direct photon flow puzzle. However, several smaller, so far neglected effects were recently recognized as possible sources of increased anisotropic flow, whereas mechanisms that tend to quench photon flow anisotropies (such as shear viscous effects) have already been accounted for. With some luck, we may be able to dispose of the photon flow puzzle by subjecting it to ``death by a thousand cuts''. 

\smallskip
\noindent
{\bf Acknowledgments: }
This work was supported by the U.S. Department of Energy under Grants No. \rm{DE-SC0004286} and (within the framework of the JET Collaboration) \rm{DE-SC0004104}. %





\begin{thebibliography}{99}

\bibitem{Owens:1986mp}
  J.~F.~Owens,
  Rev.\ Mod.\ Phys.\ 59 (1987) 465;
  S.~Catani, M.~Fontannaz, J.~P.~Guillet and E.~Pilon,
  JHEP 0205 (2002) 028;
  P.~Aurenche, M.~Fontannaz, J.-P.~Guillet, E.~Pilon and M.~Werlen,
  Phys.\ Rev.\ D73 (2006) 094007.

\bibitem{Fries:2002kt}
  R.~J.~Fries, B.~M\"uller and D.~K.~Srivastava,
  Phys.\ Rev.\ Lett.\  90 (2003) 132301;
  and 
  Phys.\ Rev.\ C72 (2005) 041902.
  
\bibitem{Zakharov:2004bi}
  B.~G.~Zakharov,
  JETP Lett.\  80 (2004) 1
   [Pisma Zh.\ Eksp.\ Teor.\ Fiz.\  80 (2004) 3].
  
\bibitem{McLerran:2014hza}
  L.~McLerran and B.~Schenke,
  arXiv:1403.7462 [hep-ph].
  
\bibitem{Bass:2002pm}
  S.~A.~Bass, B.~M\"uller and D.~K.~Srivastava,
  Phys.\ Rev.\ Lett.\  90 (2003) 082301.
  
\bibitem{Turbide:2007mi}
  S.~Turbide, C.~Gale, E.~Frodermann and U.~Heinz,
  Phys.\ Rev.\ C77 (2008) 024909.

\bibitem{Chatterjee:2013naa}
  R.~Chatterjee, H.~Holopainen, I.~Helenius, T.~Renk and K.~J.~Eskola,
  Phys.\ Rev.\ C88 (2013) 034901.
  
\bibitem{Shen:2013vja}
  C.~Shen, U.~Heinz, J.-F.~Paquet and C.~Gale,
  arXiv:1308.2440 [nucl-th].
  
\bibitem{Ruuskanen:1992hh}
  P.~V.~Ruuskanen,
  in: H. Gutbrod and J. Rafelski (Eds.), Particle production in highly excited matter,
  Vol. 303 of NATO ASI Series B: Physics, Plenum, 1992, pp. 593-613.
  
\bibitem{Adare:2008ab}
  A.~Adare, et al. [PHENIX Collaboration],
  Phys.\ Rev.\ Lett.\ 104 (2010) 132301.
  
 \bibitem{Wilde:2012wc}
  M.~Wilde [ALICE Collaboration],
  Nucl.\ Phys.\ A904-A905 (2013) 573c.
  
\bibitem{Kapusta:1991qp}
  J.~I.~Kapusta, P.~Lichard and D.~Seibert,
  Phys.\ Rev.\ D44 (1991) 2774
   [Erratum Phys. Rev D47 (1993) 4171].
   
\bibitem{Baier:1991em}
  R.~Baier, H.~Nakkagawa, A.~Niegawa and K.~Redlich,
  Z.\ Phys.\ C53 (1992) 433.
   
\bibitem{Chatterjee:2005de}
  R.~Chatterjee, E.~S.~Frodermann, U.~Heinz and D.~K.~Srivastava,
  Phys.\ Rev.\ Lett.\ 96 (2006) 202302.
  
\bibitem{Adare:2011zr}
  A.~Adare, et al.  [PHENIX Collaboration],
  Phys.\ Rev.\ Lett.\ 109 (2012) 122302.
  
\bibitem{Lohner:2012ct}
  D.~Lohner [ALICE Collaboration],
  J.\ Phys.\ Conf.\ Ser.\ 446 (2013) 012028.
  
\bibitem{Qiu:2011hf}
  Z.~Qiu, C.~Shen and U.~Heinz,
  Phys.\ Lett.\ B707 (2012) 151.
  
\bibitem{Dion:2011pp}
  M.~Dion, J.-F.~Paquet, B.~Schenke, C.~Young, S.~Jeon and C.~Gale,
  Phys.\ Rev.\ C84 (2011) 064901.
    
\bibitem{Linnyk:2013hta}
  O.~Linnyk, V.~P.~Konchakovski, W.~Cassing and E.~L.~Bratkovskaya,
  Phys.\ Rev.\ C88 (2013) 034904;
  O.~Linnyk, W.~Cassing and E.~Bratkovskaya,
  Phys.\ Rev.\ C89 (2014) 034908;
  W.~Cassing, O.~Linnyk and E.~L.~Bratkovskaya,
  arXiv:1403.5480 [nucl-th].
  
\bibitem{Gale:2012rq}
  C.~Gale, S.~Jeon, B.~Schenke, P.~Tribedy and R.~Venugopalan,
  Phys.\ Rev.\ Lett.\ 110 (2013) 012302.
  
\bibitem{Holopainen:2011pd}
  H.~Holopainen, S.~Rasanen and K.~J.~Eskola,
  Phys.\ Rev.\ C84 (2011) 064903.
  
\bibitem{Song:2007ux}
  H.~Song and U.~Heinz,
  Phys.\ Rev.\ C77 (2008) 064901.
  
\bibitem{Shen:2013cca}
  C.~Shen, U.~Heinz, J.-F.~Paquet, I.~Kozlov and C.~Gale,
  arXiv:1308.2111 [nucl-th].
  
\bibitem{Basar:2012bp}
  G.~Basar, D.~Kharzeev, D.~Kharzeev and V.~Skokov,
  Phys.\ Rev.\ Lett.\ 109 (2012) 202303.
  
\bibitem{Dusling:2009bc}
  K.~Dusling,
  Nucl.\ Phys.\ A839 (2010) 70.
  
\bibitem{vanHees:2007th}
  H.~van Hees and R.~Rapp,
  Nucl.\ Phys.\ A806 (2008) 339;
  H.~van Hees, C.~Gale and R.~Rapp,
  Phys.\ Rev.\ C {\bf 84} (2011) 054906.
  
\bibitem{Turbide:2003si}
  S.~Turbide, R.~Rapp and C.~Gale,
  Phys.\ Rev.\ C69 (2004) 014903.
  
  
\end{thebibliography}



\end{document}